\documentclass[11pt]{article}
\usepackage{moriond,epsfig}

\def\be{\begin{equation}}
\def\ee{\end{equation}}
\def\bea{\begin{eqnarray}}
\def\eea{\end{eqnarray}}

\def \Vista {{\sc Vista}}
\def \Sleuth {{\sc Sleuth}}
\def \scriptP {\ensuremath{{\cal P}}}

\def \SumPt {{\ensuremath{\sum{p_T}}}}

\def \lumi {2.0~fb$^{-1}$}

\begin{document}
\vspace*{2cm}
\title{RESULTS OF A MODEL-INDEPENDENT GLOBAL SEARCH FOR NEW PHYSICS AT CDF}

\author{CONOR HENDERSON\\for the CDF Collaboration}

\address{Massachusetts Institute of Technology,
Cambridge, MA 02139, USA.}

\maketitle\abstracts{
Data collected in Run II of the Fermilab Tevatron are searched for indications of new electroweak-scale physics.  Rather than focusing on particular new physics scenarios, CDF data are analyzed for discrepancies with the Standard Model prediction.  A model-independent approach (\Vista) considers gross features of the data, and is sensitive to new large cross-section physics.  
Further sensitivity to new physics is provided by two additional algorithms:
a ``Bump-Hunter'' searches invariant mass distributions for ``bumps'' that could indicate resonant production of new particles; and the \Sleuth\ procedure scans for data excesses in the high-\SumPt\ tails.
This combined global search for new physics in \lumi\ of $p\bar{p}$ collisions at $\sqrt{s}=1.96$~TeV reveals no indication of physics beyond the Standard Model.
}

\section{Introduction}
\label{sec:intro}

The Standard Model (SM) of particle physics has been remarkably succesful in describing observed phenomena, but is generally believed to require expansion beyond the electroweak scale.
A wide variety of possible extensions have been proposed, but no experimental evidence has yet been obtained.
As an alternative to searching for a specific new physics scenario, the present analysis attempts to perform a model-independent global search of the CDF high-$p_T$ data, to ensure that we do not miss some clear sign of new physics that happens to be lurking in an obscure place in the data.

Section~\ref{sec:vista} describes a procedure named \Vista\ which searches for discrepancies in the bulk features of the high-$p_T$ data.
An algorithm to scan invariant mass distributions for ``bumps'' that could indicate the resonant production of new particles is described in Section~\ref{sec:bump_hunter}.
Section~\ref{sec:sleuth} describes \Sleuth, a quasi-model-independent search for data excesses at high \SumPt.
Further details of this global search can be found in previous publications based on 927~pb$^{-1}$ of data.\cite{vista:prl,vista:prd}
In this global search, a discrepancy relative to the SM prediction is only considered to be significant if it corresponds to a $\geq 5\sigma$ effect.

\section{\Vista}
\label{sec:vista}

\Vista\ attempts to determine if the bulk features of the CDF high-$p_T$ data can be described in terms of the Standard Model alone.
Standard criteria are used to identify electrons, muons, taus, photons, jets, and jets tagged as originating from a $b$ quark. Significant missing transverse energy is also treated as an object. All objects are required to have $p_T \ge 17$~GeV/c.
Over four million high-$p_T$ events are considered in this \lumi\ analysis.
Events are partitioned into exclusive final states, labelled according to the objects in each state.
Standard HEP event generators 
are used to generate events from SM processes, and the response of the CDF detector is simulated by a {\sc GEANT}-based software package.

A true SM prediction, however, needs in addition some numbers that can only be obtained from the data themselves. These parameters include theoretical k-factors for the cross-sections of the SM processes, and experimental efficiencies for correctly and incorrectly reconstructing objects in the detector.
A first pass of this analysis is used to determine these correction factors from the data, and hence obtain a complete global SM prediction.

\begin{figure}
\includegraphics[width=3.25in]{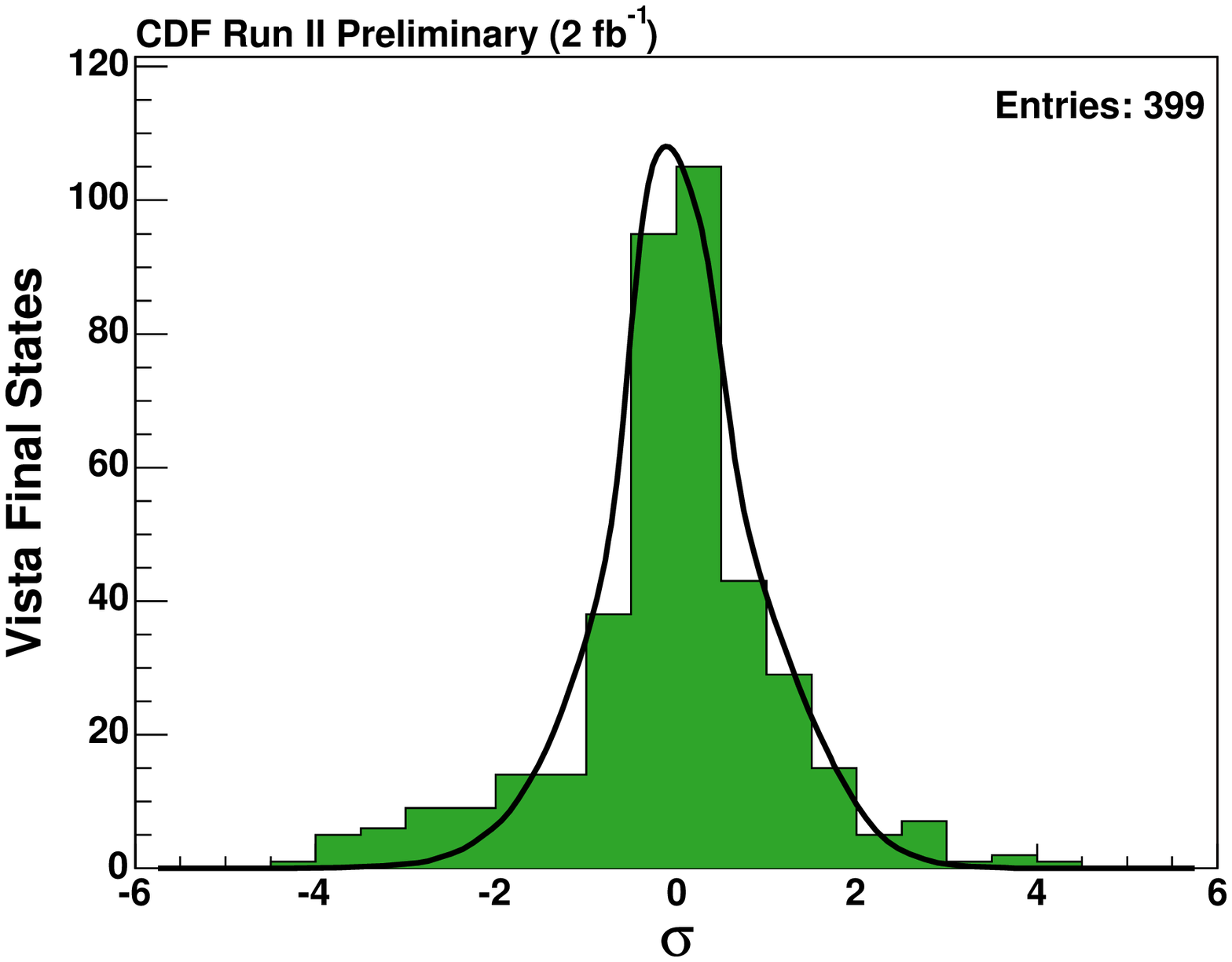}
\includegraphics[width=3.25in]{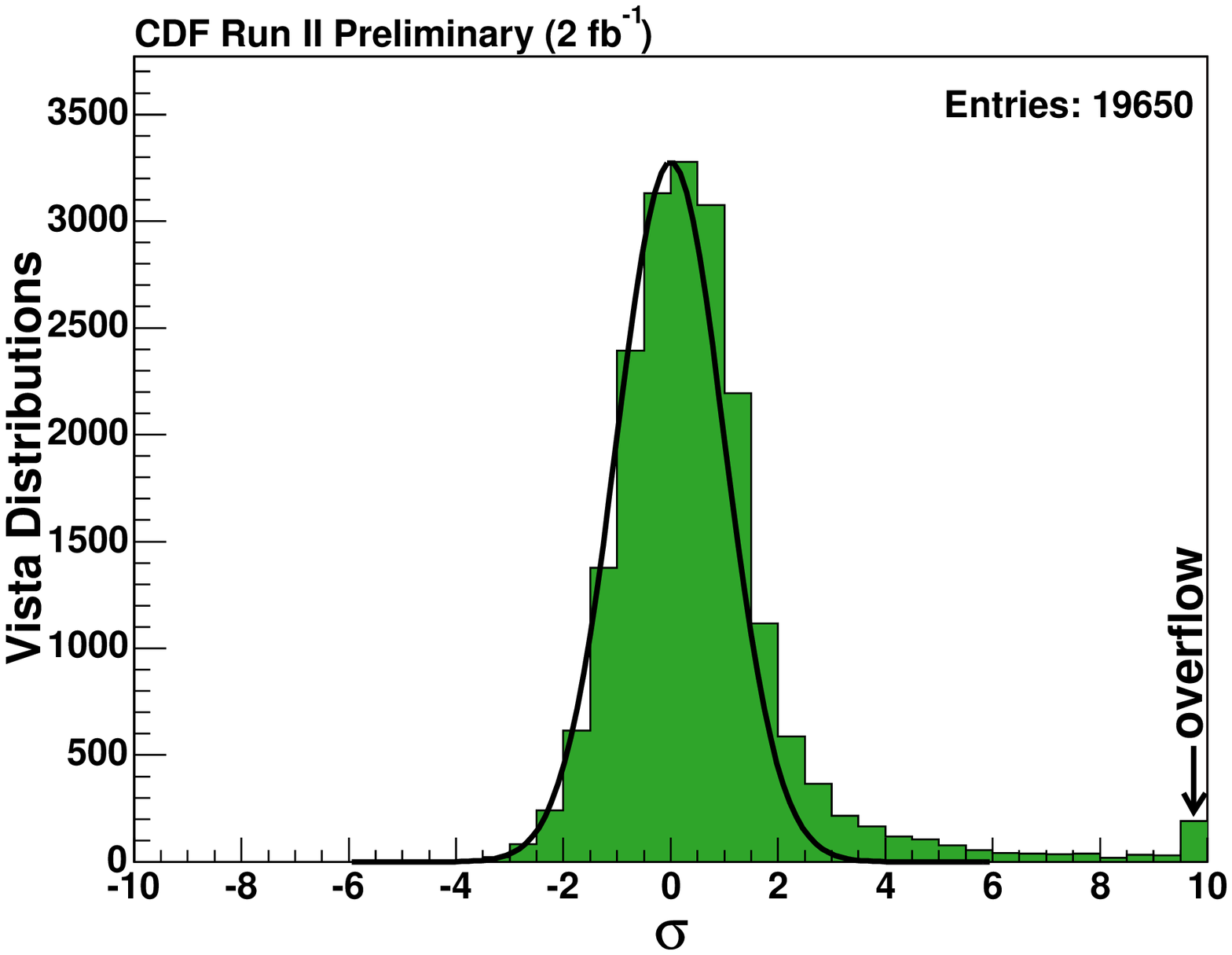}
\caption{Summary of \Vista\ global comparison of data to SM prediction for populations of final states (left) and shapes of kinematic distributions (right). The black line represents the theoretical expectation assuming no beyond-SM physics.}
\label{fig:vista_summary}
\end{figure}

The \Vista\ global comparison first studies the number of events in each of the 399 high-$p_T$ final states considered. 
The results are shown in Figure~\ref{fig:vista_summary}.
After accounting for the trials factor associated with looking at so many final states, we find that no final state exhibits a significant population discrepancy.

\begin{figure}
\includegraphics[angle=-90,width=3.25in]{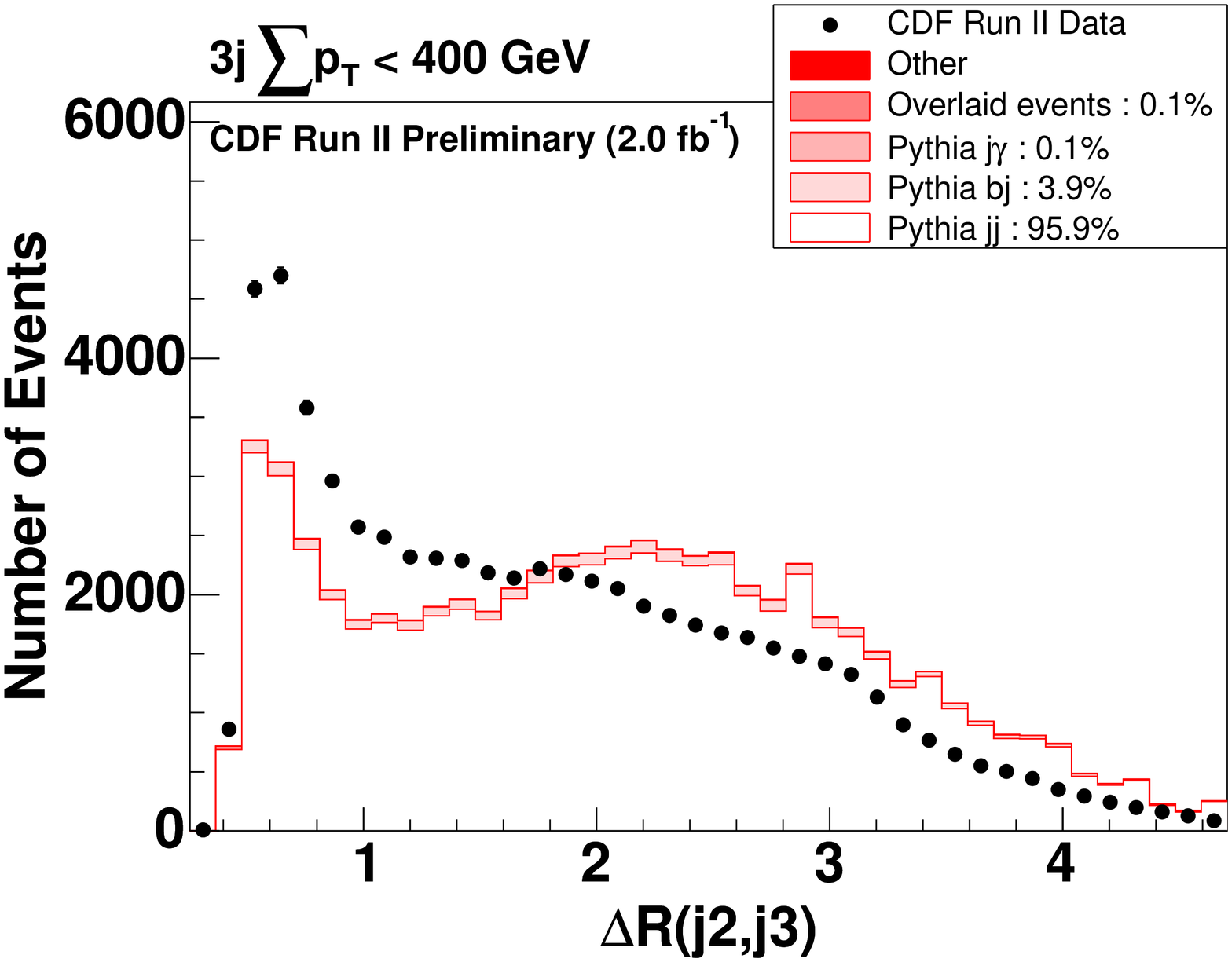}
\includegraphics[angle=-90,width=3.25in]{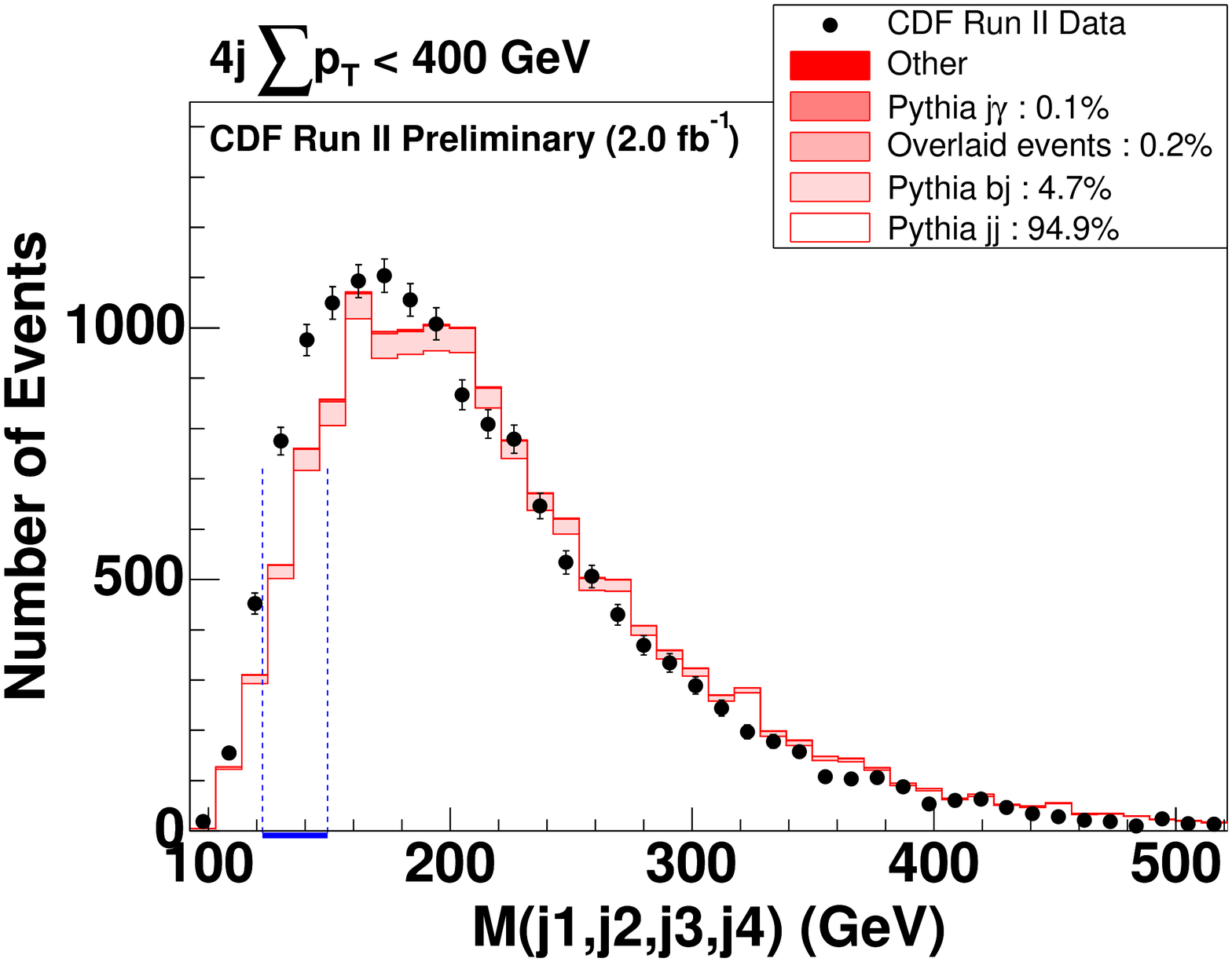}
\caption{Left: One of the significant shape discrepancies seen by \Vista, $\Delta R$ between the second and third highest $p_T$ jets in the Vista 3-jet final state.
Right: Invariant mass of all four jets in the 4-jet final state, which contains the only significant bump found by the Bump Hunter (indicated by the dashed blue lines).}
\label{fig:3j_and_bump}
\end{figure}

The shapes of kinematic distributions are also studied, using a Kolmogorov-Smirnov test to assess the agreement with the SM prediction. As shown in Figure~\ref{fig:vista_summary}, 19,650 distributions are investigated, with 559 found to have a significant discrepancy. A representative example is shown in Figure~\ref{fig:3j_and_bump}. However, analysis of these distributions reveals that they probably arise from a difficulty modelling soft QCD jet emission in the underlying Monte Carlo events. 

\section{Bump Hunter}
\label{sec:bump_hunter}

The Bump Hunter is an algorithm to search for narrow resonances in invariant mass distributions.\cite{georgios_thesis}
A search window is defined based on the expected detector resolution for the combination of objects involved. 
Each invariant mass distribution is then scanned with the appropriate sliding window, looking for bumps in the data relative to the SM background (which is taken from the \Vista\ global comparison).
The significance of any bump found is determined by pseudo-experiments, and the appropriate trials factor associated with looking in many places is accounted for.

The procedure scans 5036 invariant mass distributions, and only one is found to have a bump whose statistical significance exceeds the {\em a priori} discovery threshold.
This distribution is shown in Figure~\ref{fig:3j_and_bump}.
This bump arises from the same mismodelling of soft jets affecting the \Vista\ shapes test.

\section{\Sleuth}
\label{sec:sleuth}

\begin{figure}[htb]
\includegraphics[angle=-90,width=3.0in]{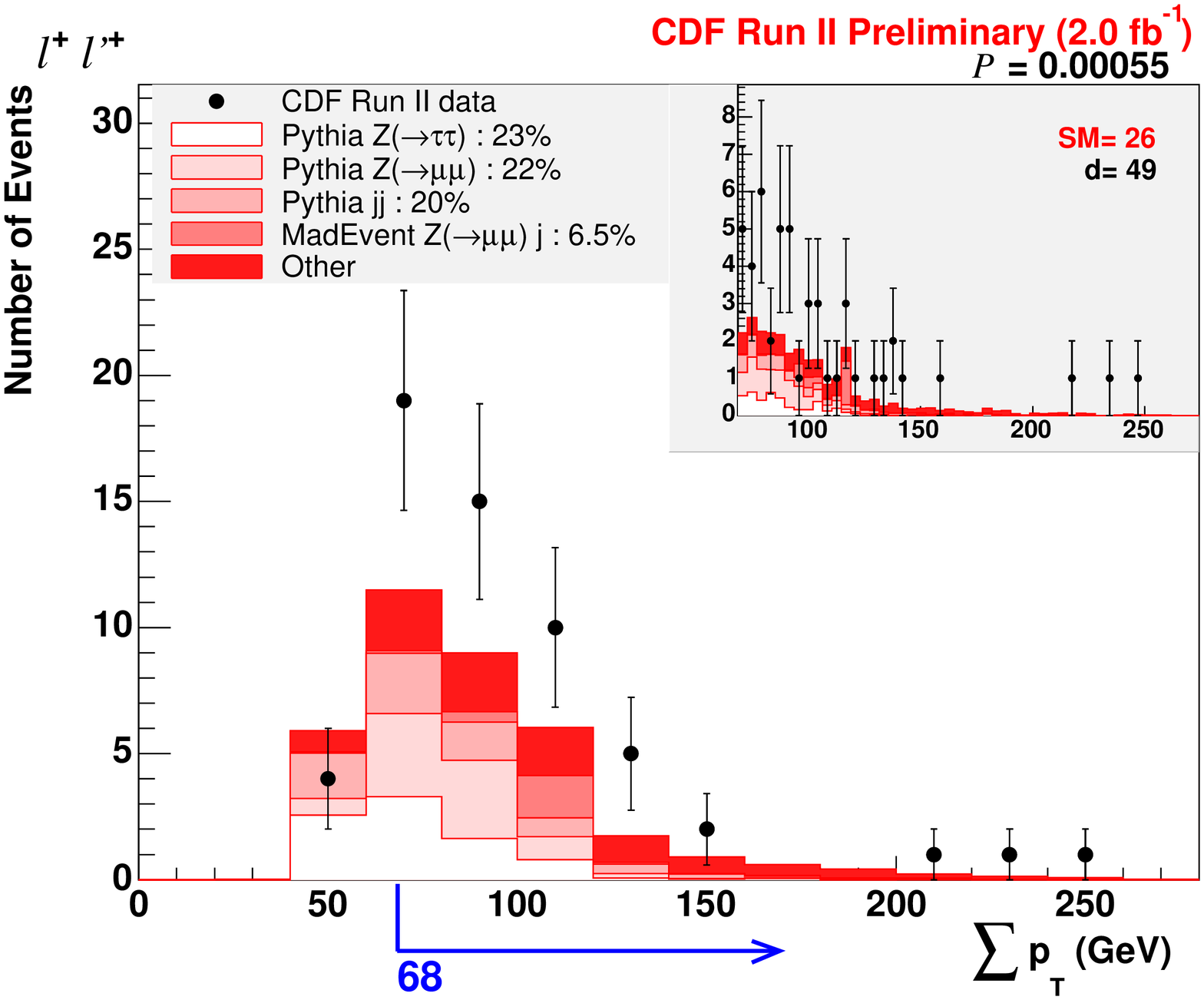}
\hfill
\includegraphics[angle=-90,width=3.0in]{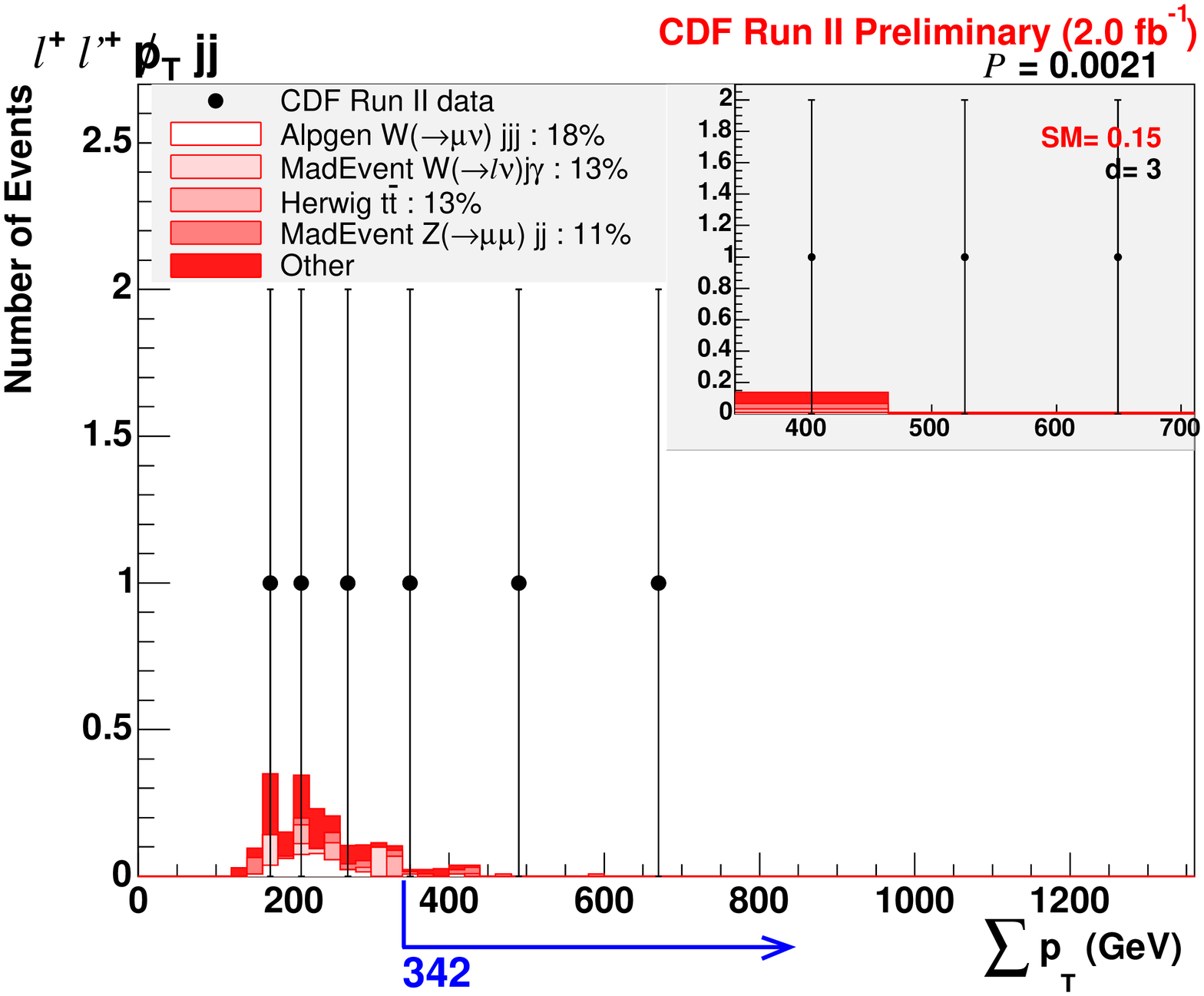}
\includegraphics[angle=-90,width=3.0in]{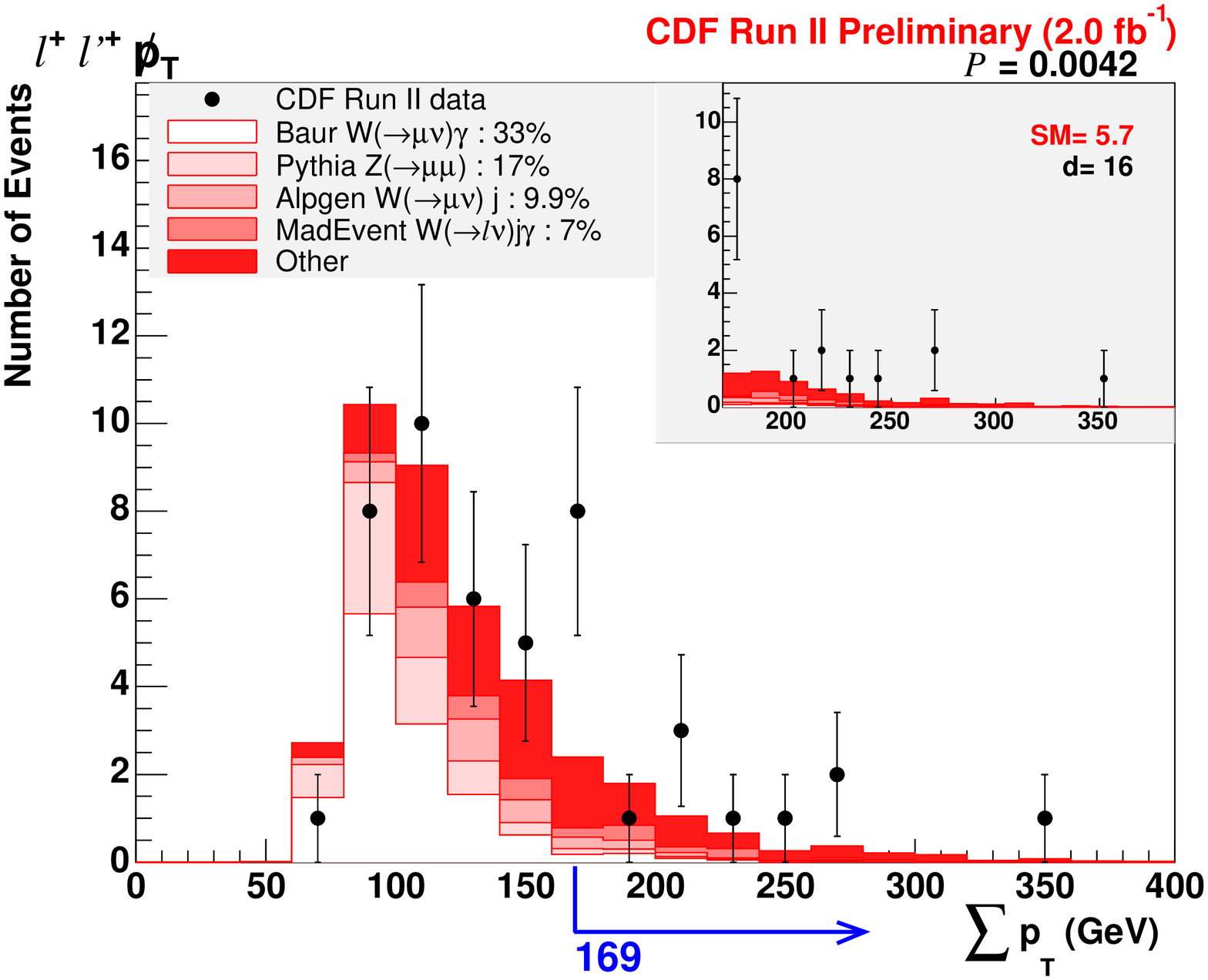}
\hfill
\includegraphics[angle=-90,width=3.0in]{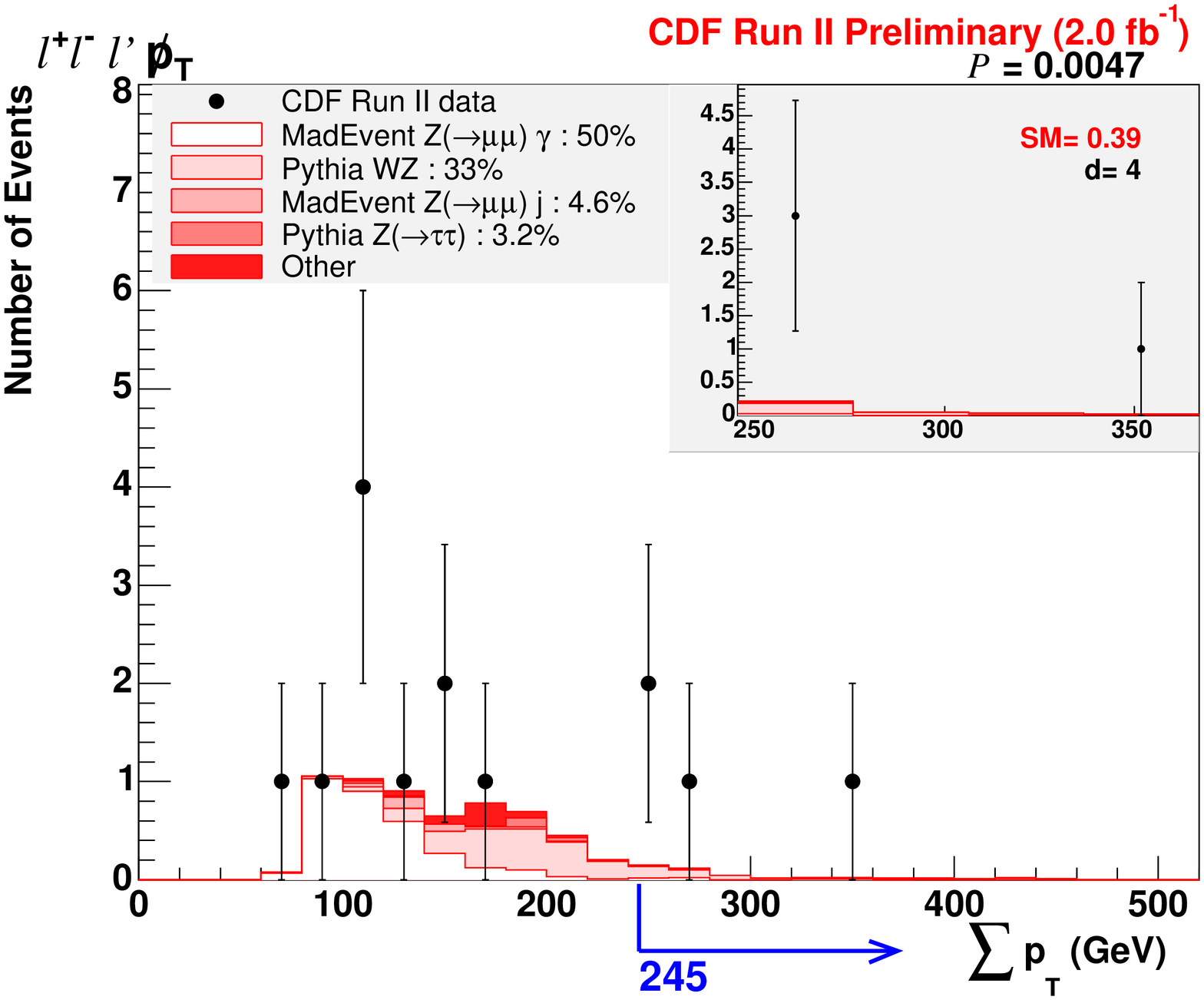}
\caption{The four most interesting final states found by \Sleuth\ in \lumi.
The label in the top left corner of each plot lists the objects in the final state, where {\em l}$^\pm$ is a lepton ($e$ or $\mu$), {\em l'} is an additional lepton of different flavor, $j$ denotes a jet and ${\not\!\!p_T}$ represents missing transverse momentum.
The region with the most significant excess of data over SM expectation is indicated by the blue line and displayed in the inset. The significance of the excess is shown by the value of \scriptP\ in the top right corner.}
\label{fig:sleuth}
\end{figure}

The \Sleuth\ algorithm considers a single variable, the summed scalar transverse momentum (\SumPt) of all objects in the event.
For each final state, \Sleuth\ determines the most interesting region on the high tail of this distribution (defined as the region with smallest probability that the integrated SM background would fluctuate up to or above the number of data events observed). The significance of the most interesting region in each final state is evaluated via pseudo-experiments, and a trials factor accounts for examining many final states.


The four most interesting final states are shown in Figure~\ref{fig:sleuth}.
While all of these final states all contain the rare signature of a same-sign electron-muon pair, none display a statistically significant excess. We estimate that $\sim8\%$ of pseudo-experiments drawn from the \Vista\ SM implementation would have produced a more significant excess in a single final state purely by chance fluctuations. Thus there is no discovery claim arising from \Sleuth\ with \lumi.

\section{Conclusions}
CDF has performed a model-independent global search for new high-$p_T$ physics in \lumi of data.
The populations of 399 exclusive final states are compared to a Standard Model prediction, but no significant discrepancy is found after accounting for the trials factor associated with looking in many places. The shapes of 19,650 kinematic distributions are also compared to the Standard Model expectations, and although 559 show a significant discrepancy, this is attributed to difficulties modelling soft QCD jet emission in the underlying Monte Carlo prediction, rather than a sign of new physics. A similar explanation is found for the results of the Bump Hunter, an algorithm to scan invariant mass distributions for narrow bumps that could indicate resonant production of new particles: only one significant bump is found, and it is attributed to the same underlying problem with QCD jet emission.
The \Sleuth\ algorithm searches the \SumPt\ spectrum of each final state, but finds no significant excesses of data over SM prediction in the tails of any single distribution.
Unfortunately, this CDF global search has not discovered new physics in \lumi.


\section*{Acknowledgments}

We are indebted to our colleagues in the CDF Collaboration, and the funding agencies which support the experiment.
We thank the Moriond QCD organizers for the invitation to attend the conference, and are grateful to
the European Union Marie Curie Series of Events for a travel grant.

\end{document}